\documentclass[aps,prl,secnumarabic,nobibnotes,twocolumn,superscriptaddress]{revtex4}
\usepackage{graphicx}
\usepackage{amsmath,bm}
\usepackage{bm}
\usepackage{natbib}
\usepackage{xcolor}
\usepackage{subfigure}
\usepackage{hyperref}

\begin{document}

\title{Nonlocal Correlation Mediated by Weyl Orbits}

\author{Zhe Hou}
\affiliation{International Center for Quantum Materials, School of Physics, Peking University, Beijing 100871, China}
\author{Qing-Feng Sun}
\email[]{sunqf@pku.edu.cn}
\affiliation{International Center for Quantum Materials, School of Physics, Peking University, Beijing 100871, China}
\affiliation{Collaborative Innovation Center of Quantum Matter, Beijing 100871, China}
\affiliation{CAS Center for Excellence in Topological Quantum Computation, University of Chinese Academy of Sciences, Beijing 100190, China}

\begin{abstract}
Nonlocality is an interesting topic in quantum physics and is usually mediated by some unique quantum states. Here we investigate a Weyl semimetal slab and find an exotic nonlocal correlation effect when placing two potential wells merely on the top and bottom surfaces. This correlation arises from the peculiar Weyl orbit in Weyl semimetals and is a consequence of the bulk-boundary correspondence in topological band theory. A giant nonlocal transport signal and a body breakdown by Weyl fermions are further uncovered, which can serve as signatures for verifying this nonlocal correlation effect experimentally. Our results extend a new member in the nonlocality family and have potential applications for designing new electric devices with fancy functions.
\end{abstract}

\maketitle{\emph{Introduction.}}
Nonlocality has always been an intriguing topic in quantum physics. The most famous one is the quantum entanglement \cite{Einstein,Horodecki} of two or more particles concerning their spin, momentum or position in the entangled state. The quantum entanglement violates the Bell inequality \cite{Bell, Gisin}, comes along with a spooky action at a distance \cite{Popescu}, and has been utilized for quantum dense encoding \cite{Bennett1} and information teleportation \cite{Bennett2}. In solid state materials, nonlocality also exists in some zero mode excitations whose properties mimic the fundamental particles in quantum field theory, such as the Majorana zero modes in a $p$-wave superconductor chain \cite{Kitaev, Fu, Lutchyn} and the Jackiw-Rebbi zero mode in a quantum spin Hall insulator constriction \cite{Klinovajia,Alicea,Wu}. Conceptually the zero modes can be regarded as half an intact particle, and similar to the positive and negative monopoles in a magnet, they have an intrinsic nonlocal correlation and can only create or annihilate in pairs.

Basically in a nonlocal phenomenon, applying a perturbation at one site of the system could arouse a remote response at the other site, which realizes a ``communication" between particles over long distances. Yet nonlocality of fermion in three dimensional bulk materials are rarely reported. The main reason stems from the electrostatic screening to the perturbation and the overall distribution of the bulk wave-functions which make the spatial response decay exponentially or proportionally to $r^{-3}$ with the distance $r$ \cite{Fetter}.

Recently a series of unconventional quantum oscillations in topological Dirac semimetals $\rm Cd_3 As_2$ were reported in experiments \cite{Moll, Zhang1, Zhang2}. In those systems, there exists a closed orbit in momentum space connecting the surface and bulk states, namely the Weyl orbit. Under the magnetic field, Weyl fermions can move along the Weyl orbit and tunnel between the top and bottom surfaces through the bulk states. The bulk and surface connection by the Weyl orbit implies that there may exist a deeper and intrinsic nonlocality in topological semimetals.

Here we report a three dimensional nonlocal correlation effect in topological Weyl semimetals (WSMs) \cite{Wan, Burkov, Xu1, Lv, Huang}. We show that by putting two attractive potential wells (PWs) merely on the top and bottom surfaces, a giant inter-surface correlation across the body would happen. The PW in the WSM works as a particle beam deflector which injects the Weyl fermions on the surface into the bulk or extract them from the bulk to the surface. With both PWs, the surface Weyl fermions would first be pumped into the bulk and then appear on the other surface, which realizes a remote nonlocal correlation between the PWs. This phenomenon arises from the unique band structure, i.e. the Weyl orbit, and is an intrinsic property in WSMs because no externally global electric or magnetic fields are applied. For Weyl fermions taking a round trip between the top/bottom surface and contributing to a constructive interference, a body breakdown by a newly formed resonant bridge state would happen, which can be a strong signature for the experimental verification on the remote correlation effect.

\maketitle{\emph{Results}}
We consider a WSM slab which is infinitely large in $x$ and $z$ directions but has finite width ($N_y$ layers) in $y$ direction, as shown in Fig. 1(a). The Hamiltonian of the WSM has a two band form \cite{Hou1,Yang,Wang}: $\hat{H}=\sum_{\bf k}c_{\bf k}^\dag H({\bf k})c_{\bf k}$ with $c_{\bf k}=(c_{\bf k \uparrow},c_{\bf k \downarrow})^T$ and $H({\bf k})=t_z \sigma_z (2+\gamma-\cos{k_x a}-\cos {k_y a} -\cos{k_z a}) \sigma_z+t_x \sigma_x  \sin {k_x a}+t_y \sigma_y \sin{k_y a}$. Here $\bf k$ is the wave-vector, $a$ is the lattice constant and $t_i  (i=x,y,z)$ is the hopping energy of nearest sites in the $i$-direction. The Pauli matrix $\sigma_i$ acts on the spin space. In this model, there exist two Weyl nodes locating at ${\bf K}_\pm=(0,0,\pm k_0 )$ with $\cos{k_0 a}=\gamma$ in the Brillouin zone. The Weyl node plays the role of magnetic monopoles with positive or negative chirality in the momentum space, and due to the bulk-boundary correspondence, the chiral surface states are formed in the WSM slab sketched in Fig. 1(a) and propagate in $\pm x$ direction in the bottom (top) surface, as shown with the yellow and green arrows. With equal energies, the surface states cannot get closed themselves in momentum space but have to transit into bulk states near the Weyl nodes, which constitute a closed loop, i.e. the Weyl orbit in Fig. 1(b).

\begin{figure}
\includegraphics[width=8.7cm, clip=]{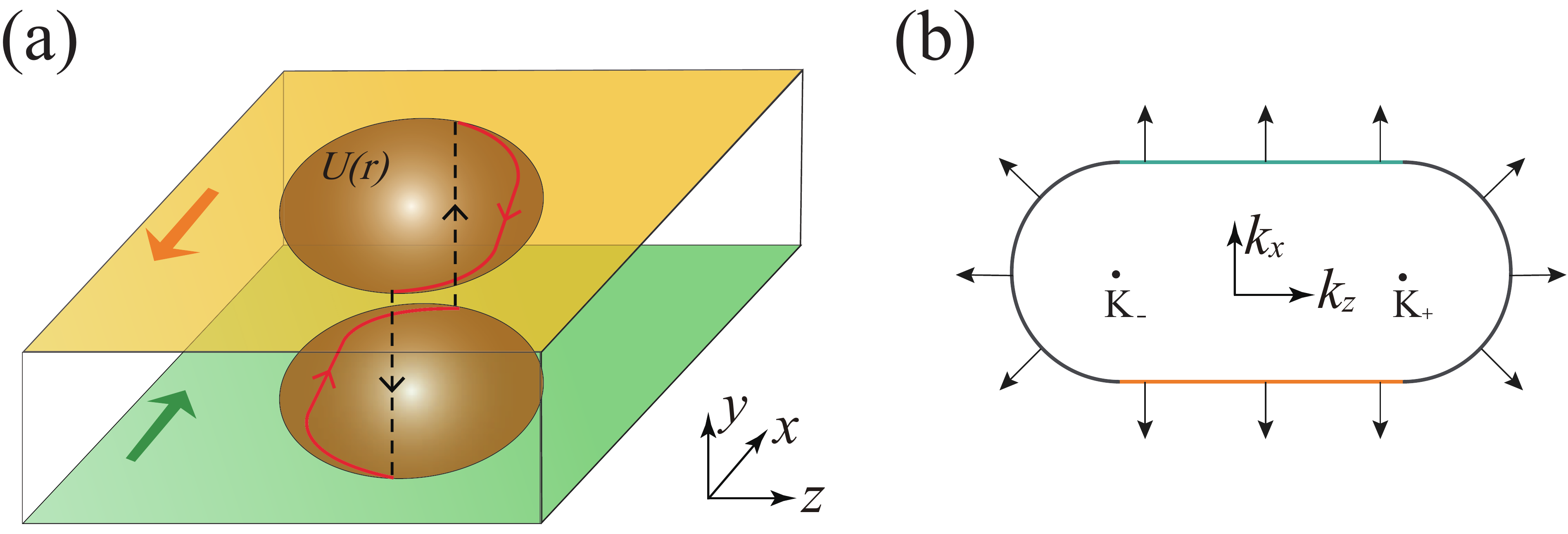}
\caption{(Color online)
Schematic diagram of the nonlocal inter-surface correlation mediated by the Weyl orbit. (a) The real space trajectory of the Weyl fermions in the WSM slab with an elliptical PW (the brown circle) on the top and bottom surfaces. The slab is infinitely large in $x$ and $z$ directions but has a thickness $N_y a$ in $y$ direction. (b) The Weyl orbit in the momentum space for the WSM slab in (a). Here the up and down line segments denote the bottom and top surface states, respectively. The black half circles denote the trivial bulk states which are connected with the surface states near the Weyl nodes. The black arrows on the Weyl orbit indicate the directions of the group velocity. }
\end{figure}

Next we show how the applied PWs induce a nonlocal inter-surface response within the Weyl orbit regime. The dark brown circles in Fig. 1(a) indicate the PWs which are on the same position of the top/bottom surfaces and yield an elliptical form: $U({\bf r})=\omega_x (x-x_C)^2+\omega_z (z-z_C )^2-\omega_x r_0^2$, if $\omega_x (x-x_C )^2+\omega_z (z-z_C )^2\leq \omega _x r_0^2$ and $y=0,N_y a$. For other positions the potential is assumed to be zero. Here $(x_C,z_C), \omega_{x,z}$ and $r_0$ are the central position, strength and size of the PW, respectively. Similar to a magnetic field, the attracting potential providing an electrostatic force ${\bf F}=-\nabla U(r)$ alters the moving trajectory of Weyl fermions and forces them into closed orbits. However, different from the electrons in normal 2D electron gas or the Dirac fermions on the surface of 3D topological insulators \cite{Xu2}, Weyl fermions can not be confined on the WSM surfaces due to the unidirectional propagation of the topological surface states (see the yellow and green segments in Fig. 1(b)).

For the Weyl fermion residing on the top surface states with a group velocity pointing in -$x$ direction, once it is trapped inside the quantum-dot like PW (see Fig. 1(a)), the attractive electrostatic force tries to change its real space trajectory (the top red curve) and meanwhile the -$z$ component pushes it into the bulk states in the momentum space (the left black half circle in Fig. 1(b)). The velocity is then tilted in the -$z$ direction and the real space trajectory is deflected to the left. The wave-function which firstly localizes on the top surface starts to permeate into the body region, and the central position of the Weyl fermion moves down as shown with the dotted black line in Fig. 1(a) (see Fig. S1 for the wave-function evolutions in the Weyl orbits) \cite{SI}.

If there exists only one PW on the top surface, the Weyl fermion is merely pumped into the body region and stops moving downward to affect the bottom surface. While if the bottom PW is placed as well, the electrostatic force would continue driving the Weyl fermion already in the body and pumps them into the bottom surface. The wave-function is then pushed from the bulk states into the bottom surface states with a group velocity in $x$ direction (the bottom red curve in Fig. 1(a)). For the Weyl fermion pumped from the bottom surface into the top one, a similar statement can be made. Thus the Weyl fermion accomplishes half a circular motion in both top/bottom surfaces under the drive of PWs. This results in a nonlocal correlation which is formed between the top and bottom PWs and mediated by the bulk states inside the Weyl orbit.

\begin{figure*}
\centering
\includegraphics[width=12cm, clip=]{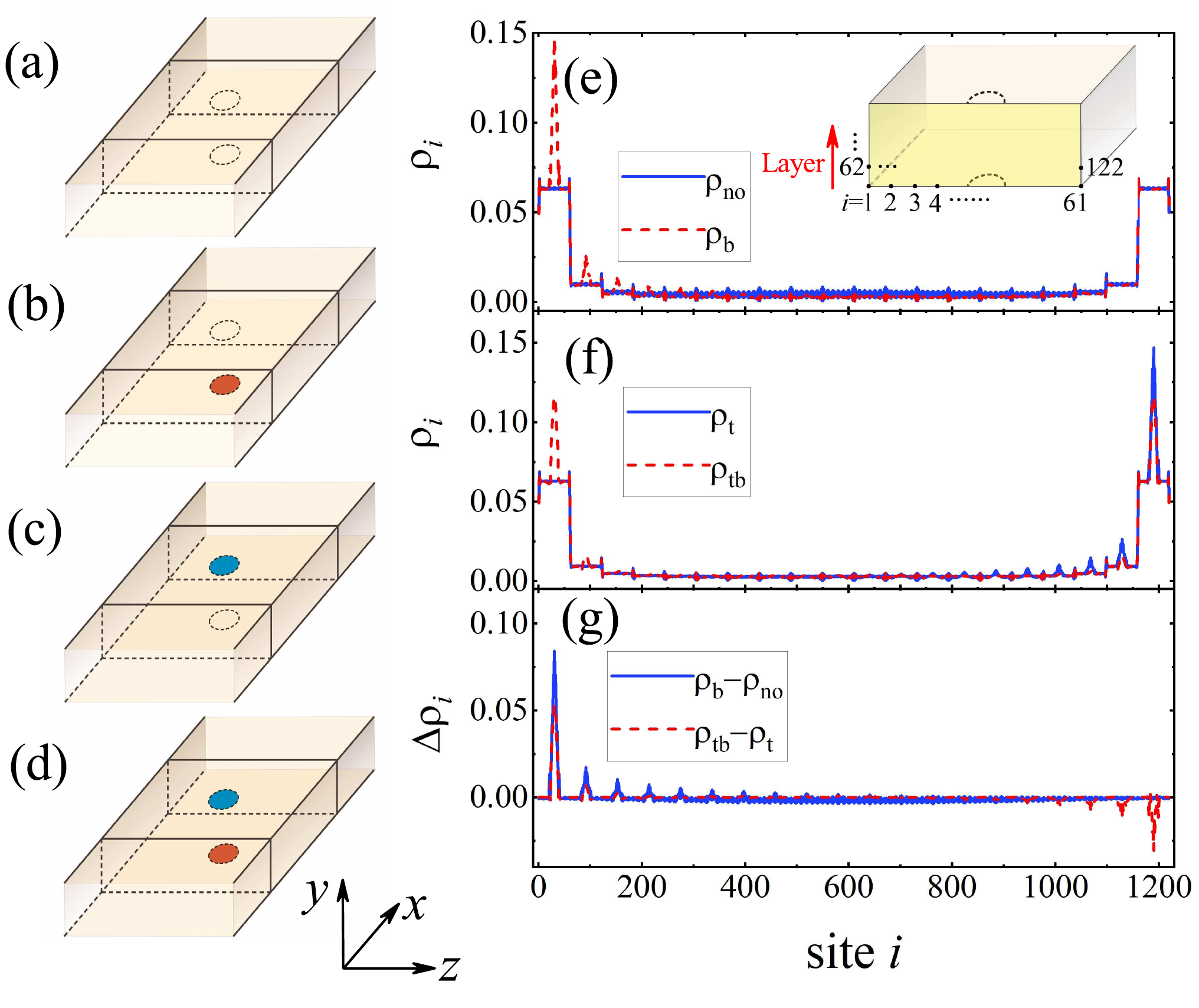}
\caption{(Color online)
Nonlocal DOS response in WSM nanoribbons. (a) to (d) Four infinitely long WSM nanoribbons (along $x$ direction) with no PWs (a), top PW (b), bottom PW (c) and top/bottom PWs (d) on the surface. (e) The DOS $\rho_{\rm no}$ and $\rho_{\rm b}$ of the center section (the bright region in the subfigure) for the nanoribbon in (a) and (b). Here the site i counts from the left to the right, and then from the bottom to the top. (f) The DOS $\rho_{\rm t}$ and $\rho_{\rm tb}$ of the center section for the nanoribbon in (c) and (d). (g) The DOS response $\rho_{\rm b}-\rho_{\rm no}$ and $\rho_{\rm tb}-\rho_{\rm t}$. Throughout the paper we set: $N_y=20,N_z=61,\gamma=0,t_x=t_y=t_z=t,r_0=10a,\omega_x=0.004t/a^2, \omega_z=0.01t/a^2$. The central position of the PWs is set on the center of the WSM slab surface. For (e-g) we set the energy $E=0.2t$.  }
\end{figure*}
To verify the nonlocal correlation discussed above, we conceive four infinitely long WSM nanoribbon systems as shown in Fig. 2(a-d). In Fig. 2(a) a pure nanoribbon with no PWs is shown, but in Fig. 2(b, c, d) the PW is put on the bottom, top and top/bottom surfaces, respectively. To eliminate that the surface states bypass the side walls and reach the other surface due to their helicity, we set the nanoribbons infinitely long in $x$ direction and restrict them in $y$ and $z$ directions with $N_y$ and $N_z$ layers. Figure 2(e) gives the DOS $\rho_i$ of the central section with and without the bottom PW (see Supplementary Material). Here the energy $E$ is chosen to be $0.2t$ above the Weyl node and the Fermi surface is a single Weyl orbit with no second bulk states mixing in. As can be seen from the no-potential case (the blue curve $\rho_{\rm no}$), the DOS dominates in the top/bottom surfaces, and decays quickly into a stable distribution away about three layers. The nonzero $\rho_i$ in the body arises from the few bulk states in the Weyl orbit. After the bottom PW is placed, the DOS $\rho_{\rm b}$ (the red dashed curve in Fig. 2(e)) shows a sharp peak in the central position of the PW and the peak height decays exponentially with the increased layer. Away from the bottom PW, the DOS $\rho_{\rm b}$ almost overlaps with $\rho_{\rm no}$, indicating that the influence of the single PW is just local \cite{Hosur}.

Figure 2(f) shows the DOS $\rho_{\rm t}$ and $\rho_{\rm tb}$ of the central section with top and top/bottom PWs, respectively. The DOS $\rho_{\rm t}$ is symmetric with $\rho_{\rm b}$ about the central site $i_c=(N_z N_y)/2+1$ because the systems in Fig. 2(b, c) can be transformed into each other by a combined operation of the time-reversal operator $\hat{T}$ and the two-fold rotation $C_2^x$ along the $x$ axis. When the bottom PW is further placed, in addition to a peak in the bottom PW, the DOS inside the top PW shows a dramatic decrease. Meanwhile the DOS of the bottom PW in turn is also influenced as compared with the peak height of $\rho_{\rm b}$ in Fig. 2(e).

In order to clearly show the nonlocal correlation, we define the DOS differences $\rho_{\rm b}-\rho_{\rm no}$ and $\rho_{\rm tb}-\rho_{\rm t}$ as the responses of the applied bottom PW and plot them in Fig. 2(g) (also see Fig. S2). While without the top PW, $\rho_{\rm b}-\rho_{\rm no}$ is almost zero when deviated from the center of the bottom PW by $3a$. However, when in the existence of the top PW, the DOS in top surface has a giant response to the bottom PW and $\rho_{\rm tb}-\rho_{\rm t}$ shows deep valley at the center of the top PW. The bottom PW itself is mutually influenced as a feedback to the top nonlocal response. The giant nonlocal response in the double PWs case indicates that there does exist a remote nonlocal correlation between the top and bottom PWs. Since the Weyl fermions can not take a side circle to reach the opposite surface, and the coupling between the top and bottom surface states is set to be weak and unable to induce such a giant nonlocal response (see Supplementary Material), the only reason that can cause this remote correlation is the Weyl orbit regime we discussed above.

Next we show that except for the DOS response, an enhanced nonlocal transport in a six-terminal system could also be induced by the nonlocal correlation. On the basis of Fig. 2(a, c, d), we additionally connect four electrodes on the top and bottom surfaces as probing terminals \cite{Hou2, Datta}. The electrode is composed of two-dimensional normal metals which can be described by a tight-binding Hamiltonian $H_{\rm NM}=\sum_{i,\sigma}E_0 a_{i\sigma}^\dag a_{i\sigma} +t_{\rm NM}\sum_{<ij>,\sigma}a_{i\sigma}^\dag a_{j\sigma}$. Here $a_{i\sigma}^\dag  (a_{i\sigma})$ is the creation (annihilation) operator at site $i$ with spin $\sigma$, $E_0$ is the on-site energy, and $<ij>$ denotes the nearest sites with a hopping amplitude $t_{\rm NM}$. Figure 3(a) shows the right side view of the six-terminal system refitted from Fig. 2(d). The top and bottom red arrows denote the topological surface states, and the yellow circles represent the PWs.

We first give the transmission coefficients $T_{m1}$ from terminal 1 to terminal  $m$ ($m$=2, 3, 4, front, and back) without the PWs in Fig. 3(b). For the incident energy $E<0.21t$, the transmission coefficients $T_{41}$ and $T_{b1}$ have large values while the other transmissions are quite small with their probability being nearly zero. This indicates that almost all the Weyl fermions propagate along the surfaces and the inter-surface transport can hardly happen. Further increasing $E$, $T_{21}$ and $T_{f1}$ have small step-like increases because the second bulk states above the Weyl orbit begin to participate.

\begin{figure}
\centering
\includegraphics[width=9cm, clip=]{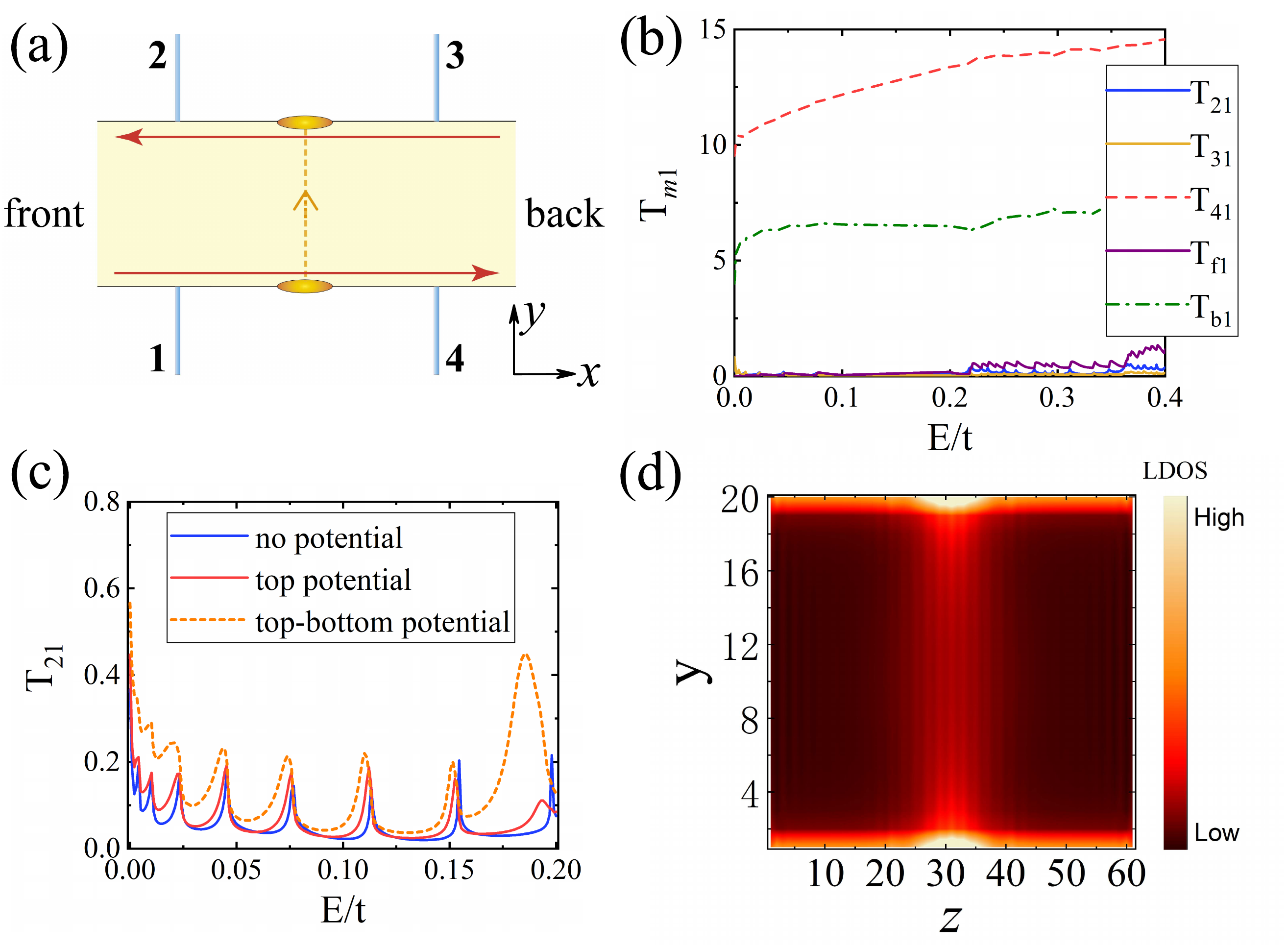}
\caption{(Color online)
Nonlocal transport in the six-terminal system. (a) The six-terminal system composed of an infinitely long WSM nanoribbon and four normal metal electrodes which are connected on the top and bottom surfaces. The red arrows denote the propagation of the surface states, and the yellow circles mark the positions of the surface PWs. The dashed line between the PWs shows the nonlocal correlation. (b) The transmission coefficients $T_{i1}$ as functions of energy $E$ in the six-terminal system without PWs. (c) The nonlocal transmission coefficient $T_{21}$ for the six-terminal system with no PWs, top PW, and top/bottom PWs, respectively. (d) The DOS profile of the central section in an infinitely long WSM nanoribbon with top/bottom PWs. The energy is set to $E=0.186t$. }
\end{figure}
In Fig. 3(c) we restrict the transport into the Weyl orbit regime by setting the energy range to $[0,0.2t]$ and investigate the influence of the surface PWs on the nonlocal transmission $T_{21}$. In the no-PWs case, $T_{21}$ has small values in most of the energy regions (see the blue curve) except for showing sharp peaks in the Van Hove singularities (VHS) of the quasi-1D sub-bands \cite{DasSarma} which arises from the finite size effect of the nanoribbon. Putting a PW on the top surface, $T_{21}$ increases slightly because few electrons in the bulk states are scattered into terminal 2. Further adding a PW on the bottom surface, the transmission coefficient $T_{21}$ shows a dramatic enhancement in the whole energy range $[0,0.2t]$ (see the dotted yellow curve) with its amplitude serval times larger than the single PW case. This enhancement can not originate from the normal scattering of the bulk states by two PWs and is a prominent sign for the nonlocal correlation between the top/bottom surfaces. Once the correlation is established by virtue of the Weyl orbit, Weyl fermions propagating in the bottom surface can be pumped into the bulk and then run into the the top surface, which constructs a transmission channel enhancing the nonlocal inter-surface transport. In Fig. S3, we design an experimental device for detecting this nonlocal transport signal.

At a critical energy $E_c=0.186t$, we note a remarkable conductance (transmission) peak with its value about 7 times of the single PW case and 13 times of the no PWs case (see Fig. S3). Although the peak position is close to the VHS of the nearest sub-band, its appearance can not be explained by the large DOS from the VHS because the height and broadening are both larger than the usual peaks of $E<E_c$. The conductance peak observed in a mesoscopic system always implies the existence of a novel quantum state. To validate this, we plot the DOS profile of the central section in the infinitely long nanoribbon at energy $E_c$ in Fig. 3(d). The bright fringes in the top and bottom layers of Fig. 3(d) indicate the surface states, and the bright spots mark the positions of the two PWs. Between the two PWs, there exists a novel bright DOS stripe with a bridge-like shape linking the top and bottom surfaces. \emph{This suggests a new resonant state formed by two spatially separated confining potentials with the wave-function spreading to the non-confined region.} We here name it the ``Weyl bridge state". Its appearance arises from the constructive interference of Weyl fermions taking a round trip between the top and bottom layers driven by the PWs. In Supplementary Material we make a semi-classical analysis on the real space trajectory of the Weyl bridge state (see Fig.S4).

In the Weyl bridge state the correlation between the PWs gets to its maximum which contributes to the large nonlocal transmission. This bridge state is quite similar to the worm-hole effect in a 3D topological insulator \cite{Rosenberg} and the 3D quantum Hall state in the topological semimetals \cite{Zhang3}. The most important distinction between them is that the Weyl bridge state does not rely on any globally applied magnetic or electric field, and is an intrinsic generation of the exotic topological band structure of WSMs. So the nonlocality discovered here should be unique in topological semimetals with a Weyl orbit structure. Finally, we want to make more discussions on the Weyl bridge state by comparing it with a natural phenomenon of lightning (Movie S1): if there accumulate enough free charges in the cloud, putting a conductor in the ground would induce an electric breakdown through the atmosphere. In our WSM systems, the top/bottom PWs play the role of the cloud and the conductor respectively, and the Weyl bridge state is just a breakdown of Weyl fermions through the WSM body.
	
\maketitle{\emph{Conclusion.}}
We have proven a nonlocal correlation effect in WSMs as a result of the Weyl orbit. As a unique property of topological semimetals, this nonlocal correlation is totally mediated by the quantum states residing on the Weyl orbit and can be simply generated by two surface PWs. The movable PWs allow a flexible and precise engineer on this nonlocal correlation and provide a new way to investigate the surface-bulk relationship in topological semimetals. The present result opens a new vision on the nonlocality in quantum physics, and has guiding significance in designing new electric device with fancy functions, such as the impurity detection and remote control in mesoscopic and micro-scales.

\maketitle{\emph{Acknowledgement}}
Z. H. thanks helpful discussions with Hua Jiang. This work was financially supported by National Key R and D Program of China (Grant No. 2017YFA0303301), NSF-China (Grant No. 11921005), the Strategic Priority Research Program of Chinese Academy of Sciences (Grant No. DB28000000), and Beijing Municipal Science \& Technology Commission (Grant No. Z181100004218001).

\end{document}